\documentclass[letterpaper]{article}
\usepackage{aaai2026}  
\usepackage{times} 
\usepackage{helvet}  
\usepackage{courier} 
\usepackage[hyphens]{url}  
\usepackage{graphicx} 
\usepackage{natbib}  
\usepackage{caption} 
\usepackage{algorithm}
\usepackage{algorithmic}
\usepackage{amsmath}
\usepackage{tikz}
\usepackage{subfig}
\usepackage{multirow}

\usepackage{newfloat}
\usepackage{listings}
\DeclareCaptionStyle{ruled}{labelfont=normalfont,labelsep=colon,strut=off} % DO NOT CHANGE THIS
\floatstyle{ruled}
\newfloat{listing}{tb}{lst}{}
\floatname{listing}{Listing}

\title{FedGuard: A Diverse-Byzantine-Robust Mechanism for Federated Learning with Major Malicious Clients}

\author{
	Haocheng Jiang\textsuperscript{\rm 1,2},
	Hua Shen\textsuperscript{\rm 1,2}\thanks{Corresponding author: cshshen@hbut.edu.cn.},
	Jixin Zhang\textsuperscript{\rm 1,2},
	Willy Susilo\textsuperscript{\rm 3},
	Mingwu Zhang\textsuperscript{\rm 1,2}\thanks{Corresponding author: mzhang@hbut.edu.cn.}
}
\affiliations{
	 \textsuperscript{\rm 1}Hubei University of Technology,
	 \textsuperscript{\rm 2}  Hubei Key Laboratory of Green Intelligent Computing Power Network\\
	\textsuperscript{\rm 3}University of Wollongong \\
	\{jianghaocheng, cshshen, zhangjx, mzhang\}@hbut.edu.cn, wsusilo@uow.edu.au
}

\usepackage{bibentry}

\begin{document}

\maketitle

\begin{abstract}
Federated learning is a distributed training framework vulnerable to Byzantine attacks, particularly when over 50\% of clients are malicious or when datasets are highly non-independent and identically distributed (non-IID). Additionally, most existing defense mechanisms are designed for specific attack types (e.g., gradient similarity-based schemes can only defend against outlier model poisoning), limiting their effectiveness. In response, we propose FedGuard, a novel federated learning mechanism. FedGuard cleverly addresses the aforementioned issues by leveraging the high sensitivity of membership inference to model bias. By requiring clients to include an additional mini-batch of server-specified data in their training, FedGuard can identify and exclude poisoned models, as their confidence in the mini-batch will drop significantly. Our comprehensive evaluation unequivocally shows that, under three highly non-IID datasets, with 90\% of clients being Byzantine and seven different types of Byzantine attacks occurring in each round, FedGuard significantly outperforms existing robust federated learning schemes in mitigating various types of Byzantine attacks.
\end{abstract}

\section{Introduction}
Federated learning operates in a distributed manner and faces a significant challenge posed by malicious clients who may alter or forge model parameters, leading to Byzantine attacks. These attacks can misguide global model training, increasing error rates, biasing outputs, and potentially causing denial-of-service issues. For example, Microsoft's chatbot Tay exhibited harmful behaviors such as racism and sexism due to learning inappropriate content \citep{SOT2018}. Similarly, language models like GPT-3, when trained on toxic data, may generate biased content \citep{JGN2025}. Byzantine attacks are classified into data poisoning attacks \citep{SMK2018, GYJ2022} and model poisoning attacks \citep{FCJ2020, XFG2025}. Data poisoning attacks alter local training data on the client side, while keeping the local model training process intact. In contrast, model poisoning attacks modify the uploaded model parameters or gradients without altering the underlying data.

Researchers have proposed various defenses. A commonly used method is robust aggregation based on gradient similarity, which aims to identify and remove outlier gradients caused by malicious clients \citep{YCK2018, CSX2017, BEE2017}. However, this type of method becomes less effective when more than half of the participants are malicious. In addition, \citeauthor{DQL2023} (\citeyear{DQL2023}) and \citeauthor{ZPA2019} (\citeyear{ZPA2019}) propose similarity attacks that actively craft malicious gradients to resemble benign ones closely. These attacks cause a high false negative rate because the system misclassifies malicious gradients as benign. Consequently, they significantly undermine the reliability of aggregation methods that rely on gradient similarity. Another commonly used approach leverages trusted data on the server side \citep{XMJ2022, MYP2024}. By maintaining a small and clean validation dataset, the server can evaluate model similarity and filter out malicious clients. Notably, \citeauthor{XMJ2022} (\citeyear{XMJ2022}) found that this type of method remains effective even when the percentage of malicious clients exceeds 50\% (up to around 60\%), making it a promising direction for further research.

However, both gradient similarity-based and trusted data defenses suffer significant degradation in effectiveness in the presence of highly non-IID data distributions or a high percentage of malicious clients. More critically, most existing Byzantine-robust strategies target only a single attack type, making them ineffective against diverse, simultaneous threats. For example, gradient similarity–based methods primarily detect outlier gradients; however, when similarity attacks and outlier attacks coexist, they fail to address both threats and may even trigger a cascading chain of misclassifications. Specifically, a misclassification of one attack type can lead to further misclassifications of other types. A typical scenario arises when Lie attacks are repeatedly misclassified and mistakenly selected. After benign clients receive a poisoned global model, the distances of their subsequently uploaded gradients become abnormally large, which in turn facilitates the success of Krum attacks, ultimately leading to the failure of the entire training process.

In summary, current Byzantine defense methods exhibit apparent limitations when facing a high percentage of malicious clients, diverse concurrent attacks, and highly non-IID data, making it difficult to ensure the robustness of federated learning in real-world applications. Specifically, several critical challenges remain unresolved:

\begin{itemize}
	\item \textbf{Attackers can easily control a large number of devices.} In practical environments, malicious adversaries often gain control over a vast number of edge devices. For instance, in certain mobile operating systems, more than 90\% of registered accounts may already be compromised \citep{YMG2019}, resulting in an extremely high percentage of malicious clients within the federated learning system.
	\item \textbf{Diverse attacks can be launched concurrently within the same round.} Malicious clients can simultaneously perform different types of attacks during a single training round, significantly weakening the effectiveness of defenses designed for specific attack types (e.g., methods relying on gradient similarity can only detect outlier Byzantine models, but are ineffective against stealthy similarity attacks).
	\item \textbf{The non-IID nature of local data further exacerbates vulnerability.} Client data often shows severe distributional heterogeneity (e.g., the data silo phenomenon \citep{LDC2022}). This characteristic amplifies the impact of Byzantine attacks, making many defense strategies more prone to failure under real-world conditions \cite{FCJ2021}.	
\end{itemize}

Creating a Byzantine-robust federated learning framework that effectively handles the three challenges simultaneously is an extremely challenging yet appealing task. To this end, we propose FedGuard, which leverages the high sensitivity of membership inference attacks to the biases exhibited during model training and their inherent vulnerabilities to achieve robust federated learning in environments characterized by highly non-IID data distributions, a high percentage of malicious clients, and diverse concurrent attacks.

The high sensitivity of membership inference attacks is evident when a model exhibits abnormal behavior. In such cases, predictions that once displayed high confidence for member samples often decrease to low confidence levels, leading to a mistaken identification of these member samples as non-members. For example, \citeauthor{ACG2016} (\citeyear{ACG2016}) demonstrated that introducing differential privacy mechanisms into gradients and model updates can significantly reduce the model's confidence in member samples, with behavior similar to Byzantine model poisoning. Furthermore, defense strategies against membership inference, such as noise injection \citep{TXS2023} and model regularization \citep{LLR2021}, indicate that membership inference attacks themselves have inherent vulnerabilities. FedGuard distinguishes the behaviors between malicious and benign clients through membership inference on training data and leveraging these inherent vulnerabilities, thereby defending against Byzantine attacks. The major contributions of this work are as follows:

\begin{enumerate}
	\item We design a Byzantine defense method using membership inference. While Byzantine attacks can weaken this approach, our detection method effectively distinguishes between benign and malicious models by analyzing the inference differences between clients. It focuses on the behavioral patterns of malicious clients without relying on the features of a specific attack or strict data distribution assumptions. This enables it to identify abnormal model behaviors even in complex scenarios with a high percentage of malicious clients, diverse attack types, and severely non-IID client data, demonstrating strong generality.
	\item We introduce FedGuard, a federated learning mechanism specifically designed to protect against a significant number of malicious clients and various Byzantine attacks. FedGuard combines our designed defense method with federated learning by incorporating shadow models that simulate client behaviors, along with pre-training a defense model. The server utilizes these shadow models to simulate both benign and malicious behaviors, and it pre-trains the defense model based on their confidence scores. Throughout the federated learning process, this defense model is capable of identifying and isolating malicious clients in real time.
	\item We conducted extensive experiments to verify the robustness of FedGuard. Experimental results on three representative datasets demonstrate that FedGuard remains robust even under highly non-IID data, with a percentage of malicious clients as high as 90\% and seven concurrent Byzantine attacks present in each round. Under these extreme conditions, the performance of FedGuard consistently outperforms existing Byzantine-resistant federated learning methods.
\end{enumerate}

\section{Background and Related Work}
\subsection{Federated Learning}
To leverage distributed data efficiently,  \citeauthor{MMR2017} (\citeyear{MMR2017}) introduced the concept of Federated Learning (FL). A typical FL process consists of three steps: 1) The server distributes the current global model to all participating clients; 2) Each client performs local training on its private dataset based on the received global model, then uploads the resulting local model update to the server; 3) The server aggregates the received updates from all clients using a predefined aggregation rule to generate a new global model. This process is iteratively repeated until a predefined convergence or termination condition is met.

Federated learning has been widely applied in practice. Google was one of the pioneers in using federated learning for the next-word prediction model in its Gboard keyboard \citep{AKR2019}, and further integrated differential privacy techniques for deployment in real-world production \citep{ZYG2023}. Similarly, Apple also utilized federated learning combined with differential privacy to train virtual assistants like Siri \citep{ACN2025}. Federated learning has shown great potential in distributed training. However, research by \citeauthor{BEE2017} (\citeyear{BEE2017}) pointed out that malicious attacks, such as software failures or data poisoning by clients, can lead to incorrect model updates being selected by the server, causing the model to train in the wrong direction. This issue, known as Byzantine attacks in federated learning, has become a significant challenge hindering its further development.

\subsection{Byzantine-robust Federated Learning}
To achieve robust aggregation in federated learning,  \citeauthor{BEE2017} (\citeyear{BEE2017}) proposed the Krum and Multi-Krum aggregation rules. Krum aggregates gradients by selecting the one most similar to others, while Multi-Krum further selects multiple similar gradients to enhance robustness. Later,  \citeauthor{EEG2018} (\citeyear{EEG2018}) introduced Bulyan, which first filters gradients using existing Byzantine-robust aggregation rules and then applies median aggregation to further reduce the impact of malicious workers. In addition, \citeauthor{YCK2018} (\citeyear{YCK2018}) proposed the Median and Trimmed-mean aggregation rules. The Trimmed-mean rule computes the average after removing extreme values, while the Median rule updates the model using the median of the model parameters from each client. However, these methods can only defend against Byzantine attacks from fewer than half of the clients. Furthermore, \citeauthor{FCJ2020} (\citeyear{FCJ2020}) proposed a model poisoning attack, in which carefully altered local model parameters enable the attack model to bypass the above defense strategies effectively.

In response to this more challenging type of attack, researchers have proposed robust aggregation strategies based on trusted datasets. For example, \citeauthor{XMJ2022} (\citeyear{XMJ2022}) introduced the FLTrust method, where the server maintains a small trusted dataset and assigns trust scores based on the similarity in direction between the model updates uploaded by clients and the server's own model update. This method can effectively resist attacks and ensure the performance of the global model even when the proportion of malicious clients exceeds half. \citeauthor{MYP2024} (\citeyear{MYP2024}) used a global model generated from a root dataset as a trusted standard to assess the anomaly of client model updates. Additionally, \citeauthor{SHY2022} (\citeyear{SHY2022}) proposed the DiverseFL method, which uploads a small number of representative data samples from each client to the server's trusted execution environment (TEE) in advance, using these samples to guide the training of updates. \citeauthor{YSL2024} (\citeyear{YSL2024}) further proposed a method where a small-scale clean dataset maintained on the server is used to train a guiding model, which compares with client updates to detect abnormal updates. Recently, \citeauthor{JFL2025} (\citeyear{JFL2025}) introduced the FedSV method, which incorporates a self-validation mechanism, using a small amount of clean validation data maintained by the server to evaluate the performance of the models uploaded by clients, selecting those with positive effects for aggregation. Compared to these trusted data-based methods, FedGuard requires a smaller amount of trusted data and is more feasible in practical application scenarios.

Methods based on trusted data (such as FLTrust) have shown certain potential in scenarios where most clients are Byzantine nodes. Therefore, we chose to utilize trusted data to design FedGuard. Note that FedGuard differs from existing trusted data-based FLs in how trusted data is used to resist Byzantine attacks. This approach is one of our core innovations, enabling FedGuard to maintain robustness and stability even in situations with highly non-IID data, a very high percentage of malicious clients, and varying types of Byzantine attacks launched by malicious clients in each training round—something that existing trusted data-based FL methods cannot achieve. Furthermore, compared with existing trusted data-based methods, FedGuard requires a significantly smaller amount of trusted data, thereby enhancing its practicality and deployability in real-world scenarios.

\section{FedGuard}

\subsection{Overview of FedGuard}
In FedGuard, malicious clients can launch concurrent attacks in various forms and even access the models of other benign clients. In addition, the server is regarded as a trusted entity, responsible for maintaining a clean public seed dataset, which can be obtained through manual labeling \cite{DCL2021} or constructed from open-source datasets.

FedGuard exploits vulnerabilities in membership inference, where Byzantine attacks lessen the model's bias towards training data. This results in differing membership inference outcomes for benign and malicious clients, allowing for effective differentiation between them. FedGuard integrates this method with federated learning and operates in two phases: offline and online.

The offline phase only involves the server and no real clients. The server begins by creating a public dataset and generating shadow models that simulate benign and malicious client behaviors. It then randomly selects one benign shadow model as the reference model, which serves as a benchmark for later stages. Using the shadow models, reference model, and public dataset, the server obtains a defense model to infer the behaviors of real clients in the online phase. The online phase refers to the stage where all clients have gone online and started collaborative model training with the server. Initially, each client creates the same dataset used by the server during the offline phase. Clients then train local models using this public dataset and their private dataset. Simultaneously, the server employs the pre-trained defense model to perform real-time discrimination on client models, identifying and filtering potential Byzantine clients to prevent malicious models from contaminating the global model.

Note that, unlike traditional membership inference methods \citep{SSS2017}, which primarily rely on the outputs of shadow models for discrimination, our approach distinguishes clients by comparing the output differences between shadow models and the reference model. Specifically, we use the mean squared error (MSE) of prediction confidences and the true-class confidence difference (TCD) as discriminative features, as these two features better capture the behavioral discrepancies between the two types of models.

\subsection{Details of FedGuard}

\subsubsection{Offline phase}
\begin{figure}[htbp]
	\centering
	\includegraphics[width=1.0\linewidth]{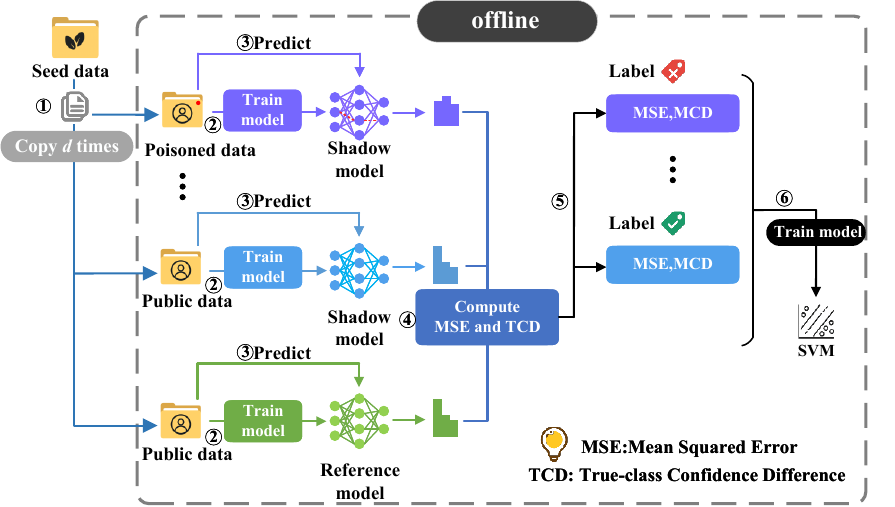}
	\caption{The Offline Phase of FedGuard}
	\label{fig:FedGuard Offline phase}
\end{figure}

As shown in Figure \ref{fig:FedGuard Offline phase}, the server first collects a small set of strictly clean samples as the seed dataset $D_{\text{seed}}$. To enhance the model's sensitivity during subsequent inference, the server replicates this dataset $d$ times, constructing a public dataset $D_{\text{pub}}$. This replication amplifies the confidence differences of the model on the training samples, making Byzantine behaviors easier to detect.

Next, the server trains shadow models to simulate the update behaviors of different clients, generating samples with identity labels for the subsequent defense model training. The server trains models on $D_{\text{pub}}$ to create shadow models for benign clients that simulate their behaviors. For malicious clients, it applies known Byzantine attack strategies (e.g., data poisoning and model poisoning) to generate shadow models that simulate their behaviors on $D_{\text{pub}}$. The server then randomly selects a shadow model that simulates a benign client to be a reference model, which serves as the benchmark for comparison in subsequent stages.

After that, the server extracts discriminative features from the prediction differences between each shadow model $f_{\text{shadow}}$ and the reference model $f_{\text{ref}}$ on $D_{\text{pub}}$ for training a defense model $f_{\text{def}}$. It feeds $D_{\text{pub}}$ into both $f_{\text{shadow}}$ and $f_{\text{ref}}$ to obtain their respective confidence scores $\mathbf{C}_{\text{shadow}}$ and $\mathbf{C}_{\text{ref}}$. Suppose $D_{\text{pub}}$ contains $L$ classes and $N$ samples. $\mathbf{C}_{\text{shadow}}$ and $\mathbf{C}_{\text{ref}}$ are both vectors of length $N$, where the $i$th component $\mathbf{C}^i$ represents the confidence score of the $i$th sample. For the $i$th sample of $D_{\text{pub}}$ with the true class label $l_i$, the following two metrics are computed:

$$\mathrm{MSE}=\frac{1}{NL}\sum_{i=1}^N\left\|\mathbf{C}_\mathrm{shadow}^i-\mathbf{C}_\mathrm{ref}^i\right\|_2^2, $$

$$\mathrm{TCD}=\frac{1}{N}\sum_{i=1}^N\left|\mathbf{C}_{\mathrm{shadow}}^i[l_i]-\mathbf{C}_{\mathrm{ref}}^i[l_i]\right|.$$

\noindent MSE quantifies the deviation between $f_{\text{shadow}}$ and $f_{\text{ref}}$, while TCD assesses the confidence difference for the true class. These features form the vector \(\mathbf{x} = [\text{MSE}, \text{TCD}]\). Suppose the server creates $k$ shadow models in the offline phase. The identity label of a shadow model is \(y \in \{0, 1\}\) (0 for malicious and 1 for benign). The set \(\{(\mathbf{x}_s, y_s)\}\) ($s$ $=$ $1$, $\cdots$, $k$) is used as the training set $D_{\text{def}}$ for training the defense model \(f_{\text{def}}\). A Support Vector Machine is used for this binary classification, as illustrated in Algorithm \ref{alg: Defense Model Training}.

\begin{algorithm}[tb]
\caption{Defense Model Training}
\label{alg: Defense Model Training}
\textbf{Input}: $N$, $L$, real label $l$, $\mathbf{C}_\text{shadow}$, $\mathbf{C}_\text{ref}$, $k$, regularization parameter $\lambda$, learning rate $\eta$, maximum iterations $T$ \\
\textbf{Output}: Weight vector $\mathbf{w}$, bias $b$
\begin{algorithmic}[1]
	\STATE Initialize $\mathrm{sum\_mse} \gets 0$, $\mathrm{sum\_tcd} \gets 0$, $D_\text{def} \gets \emptyset$
	\FOR{$s=1$ to $k$}
	\STATE $\mathrm{sum\_mse} \gets 0, \ \mathrm{sum\_tcd} \gets 0$
	\FOR{$i=1$ to $N$}
	\STATE $\mathbf{diff\_vec} \leftarrow \mathbf{C}_{\mathrm{shadow}}^{i} - \mathbf{C}_{\mathrm{ref}}^{i}$
	\STATE $\mathrm{sum\_mse} \gets \mathrm{sum\_mse} + \left\| \mathbf{diff\_vec} \right\|_2^2$
	\STATE $\mathrm{true\_diff} \leftarrow \left| \mathbf{C}_{\mathrm{shadow}}^{i}[l_i] - \mathbf{C}_{\mathrm{ref}}^{i}[l_i] \right|$
	\ENDFOR
	\STATE $\mathrm{MSE}_s \leftarrow \frac{\mathrm{sum\_mse}}{N \times L}$, $\mathrm{TCD}_s \leftarrow \frac{\mathrm{sum\_tcd}}{N}$
	\STATE $\mathbf{x} \leftarrow (\mathrm{MSE}_s, \mathrm{TCD}_s)$
	\STATE $y \leftarrow \begin{cases} 0 & \text{Data Poisoning or Model Poisoning} \\ 1 & \text{otherwise} \end{cases}$
	\STATE $D_\text{def} \gets D_\text{def} \cup \{(\mathbf{x}, y)\}$
	\ENDFOR
	
	\STATE Initialize $\mathbf{w} \gets \mathbf{0}$, $b \gets 0$
	\FOR{$t=1$ to $T$}
	\FOR{each $(\mathbf{x}, y) \in D_\text{def}$}
	\IF{$y (\mathbf{w}^\mathrm{T} \mathbf{x} + b) < 1$}
	\STATE $\mathbf{w} \gets \mathbf{w} - \eta (\lambda \mathbf{w} - y \mathbf{x})$
	\STATE $b \gets b + \eta y$
	\ELSE
	\STATE $\mathbf{w} \gets \mathbf{w} - \eta \lambda \mathbf{w}$
	\ENDIF
	\ENDFOR
	\ENDFOR
\STATE \textbf{return} $\mathbf{w}, b$
\end{algorithmic}
\end{algorithm}

\subsubsection{Online phase}
\begin{figure}[ht]
	\centering
	\includegraphics[width=1\linewidth]{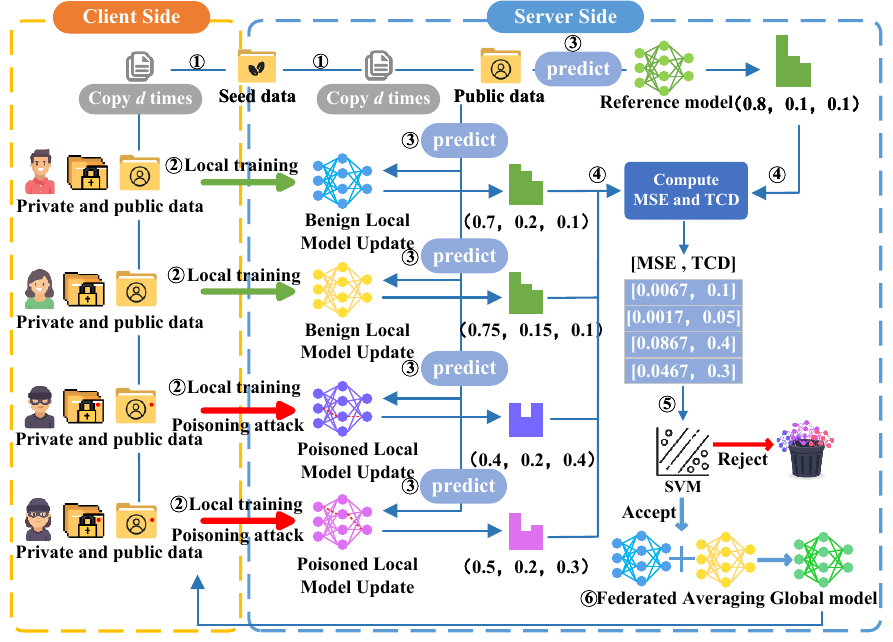}
	\caption{The Online Phase of FedGuard}
	\label{fig: FedGuard Online phase}
\end{figure}

As shown in Figure \ref{fig: FedGuard Online phase}, during the training initialization, the server distributes the seed dataset $D_{\text{seed}}$ to each client, who then replicates it $d$ times to construct the public dataset $D_{\text{pub}}$. At the beginning of each training round, the server broadcasts the global model. Clients perform local training using $D_{\text{pub}}$ and their private dataset $D_{\text{pri}}$ before uploading their model updates. For each model uploaded by the clients, the server feeds $D_{\text{pub}}$ into it to obtain the predicted confidence $\mathbf{C}_{\text{client}}$. The server compares $\mathbf{C}_{\text{client}}$ with $\mathbf{C}_{\text{ref}}$ to calculate the corresponding MSE and TCD features for the corresponding client. Then, the server feeds this client's $[ \text{MSE}, \text{TCD} ]$ features into the pre-trained defense model $f_{\text{def}}$ to determine whether the client is malicious. In extreme cases, if all clients are considered malicious in a round, the client with the smallest MSE is selected as benign. Only the models of clients classified as benign will participate in the current round of global aggregation, thus mitigating the impact of malicious updates.

\section{Experiments}
We conduct experiments using PyTorch on a machine with an Intel Xeon E5-2696 v2 processor (24 cores), an NVIDIA V100 GPU (16 GB), and 64 GB of RAM.

\subsection{Datasets and Models}

\textbf{Datasets:} To evaluate the robustness of our FedGuard, we selected three benchmark datasets covering diverse tasks.
\begin{itemize}
	\item \textbf{MNIST} \cite{LBB1998}: A benchmark for handwritten digit classification, consisting of 70,000 28$\times$28 grayscale images across 10 digit classes (0–9).
	\item \textbf{Fashion-MNIST} \cite{HKR2017}: A benchmark for clothing classification, featuring 70,000 28$\times$28 grayscale images across 10 classes.
	\item \textbf{GTSRB} \cite{SSS2011}: A dataset for traffic sign recognition with 43 classes, containing 39,209 training images and 12,630 test images in RGB format.
\end{itemize}

\textbf{Dataset Partitioning:} In the IID scenario, samples are uniformly distributed among clients. In the non-IID scenario, we use a Dirichlet distribution ($\alpha$) for partitioning. For each class, samples are allocated to clients based on $\alpha$, where $\alpha$ controls the level of heterogeneity. A smaller $\alpha$ increases the discrepancy in client distributions, enhancing non-IID characteristics, while a larger $\alpha$ results in more balanced distributions.

\textbf{Models:} For the MNIST and Fashion-MNIST datasets, we trained a convolutional neural network with the following architecture: Conv2d($1$, $32$, $3$, $padding=1$) $\rightarrow$ MaxPool($2\times2$) $\rightarrow$ Conv2d($32$, $64$, $3$, $padding=1$) $\rightarrow$ MaxPool($2\times2$) $\rightarrow$ Flatten $\rightarrow$ Linear($64\times7\times7$, $128$) $\rightarrow$ Linear(128, 10) $\rightarrow$ Softmax. For the GTSRB dataset, we used a deeper convolutional neural network: Conv2d($3$, $64$, $3$, $padding=1$) $\rightarrow$ MaxPool($2\times2$) $\rightarrow$ Conv2d($64$, $128$, $3$, $padding=1$) $\rightarrow$ MaxPool($2\times2$) $\rightarrow$ Conv2d($128$, $128$, $3$, $padding=1$) $\rightarrow$ MaxPool($2\times2$) $\rightarrow$ Flatten $\rightarrow$ Linear($128\times4\times 4$, $256$) $\rightarrow$ Linear($256$, $43$) $\rightarrow$ Softmax.

\subsection{Evaluated Poisoning Attacks}
We use the following data and model poisoning attacks in our experiments.

\textbf{Data Poisoning Attacks:} 1) Label-Flipping Attack. Byzantine clients randomly replace a portion of true labels with incorrect ones to mislead the global model. 2) Backdoor Injection \cite{TBS2019,YXG2025}. Byzantine clients inject samples with specific backdoor triggers into their local data, such as setting certain pixels to fixed values in image tasks.

\textbf{Model Poisoning Attacks:} 1) Sign-Flip Attack. Byzantine clients invert each parameter of their local model updates. 2) Random Parameters Attack. Byzantine clients replace local model parameters with random values. 3) Bit-Flip Attack. Byzantine clients flip a random or specified number of bits in the binary representation of model parameters. 4) Lie Attack \cite{BBG2019}. Byzantine clients add small-magnitude Gaussian noise to local model parameters. 5) Krum Attack \cite{FCJ2020}. Byzantine clients scale down local parameters to bypass robust aggregation algorithms.

\subsection{Baseline Aggregation Rules}
Bulyan and Loss Function-based Rejection (LFR) assume the number of malicious clients is known. Let $n$ be the total number of participating clients in a given round, and $f$ the estimated number of Byzantine clients. We adopted the following aggregation rules as benchmarks: 1) \textbf{FedAvg} \cite{MMR2017}. Aggregates local model parameters from clients by averaging them to update the global model. 2) \textbf{Bulyan} \cite{EEG2018}. Uses the Krum aggregation rule to select the $n-2f$ most reliable updates, choosing the closest value to the median for each parameter dimension and averaging them for the final update. If the number of Byzantine clients exceeds half the total, it selects $n-f$ reliable updates instead. 3) \textbf{Median} \cite{CD2023}. Computes the coordinate-wise median of all client updates as the global model parameters. 4) \textbf{LFR} \cite{FCJ2020}. Evaluates each local update on a clean validation dataset, selecting the $n-f$ updates that reduce validation loss the most. These updates are then averaged for the global model. 5) \textbf{FLTrust} \cite{XMJ2022}. Uses a server-side model, trained on a clean dataset, to compute cosine similarity between client updates and the server update. Only trusted updates are selected, normalized, and aggregated with weights proportional to their similarity.

\subsection{SVM Classification Performance}
\begin{figure}[htbp]
	\centering
	\subfloat[MNIST]{%
		\includegraphics[width=0.32\linewidth]{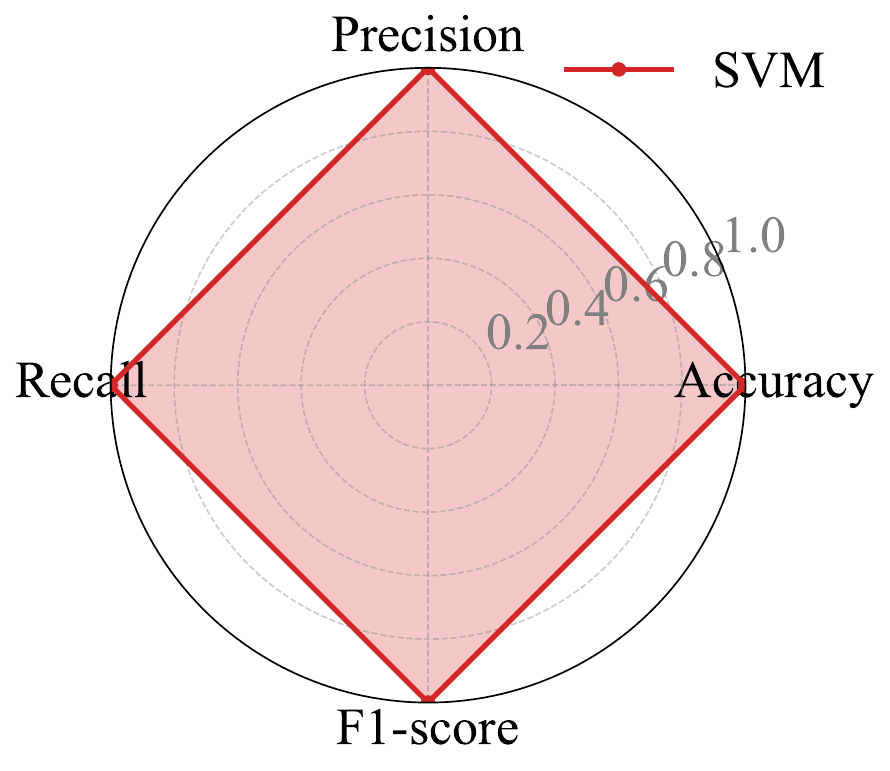}
		\label{fig:subA}}
	\hfill
	\subfloat[FashionMNIST]{%
		\includegraphics[width=0.32\linewidth]{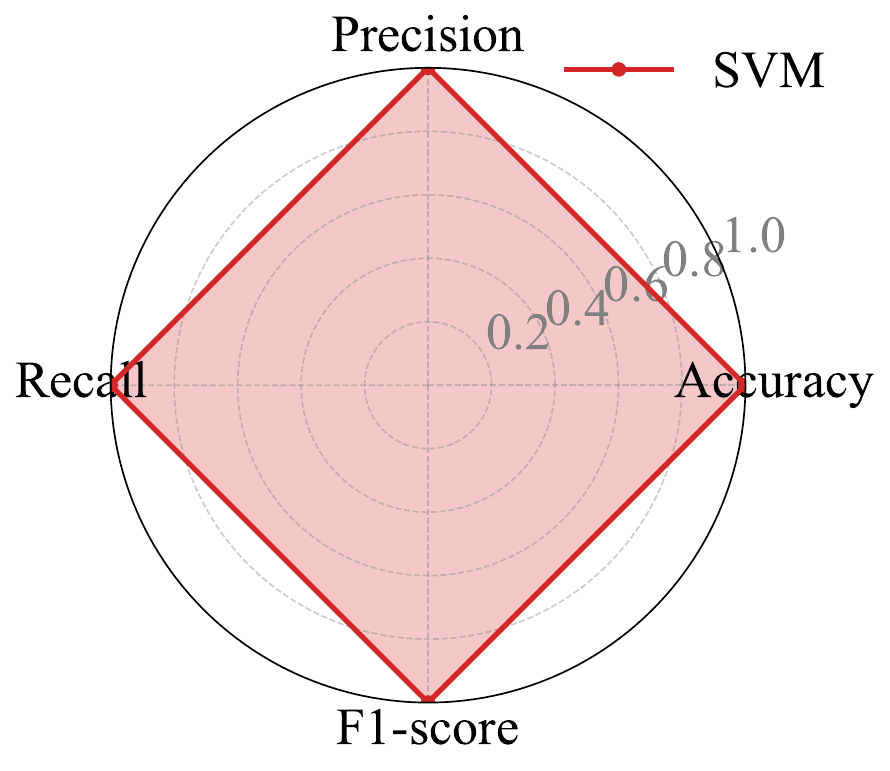}
		\label{fig:subB}}
	\subfloat[GTSRB]{%
		\includegraphics[width=0.32\linewidth]{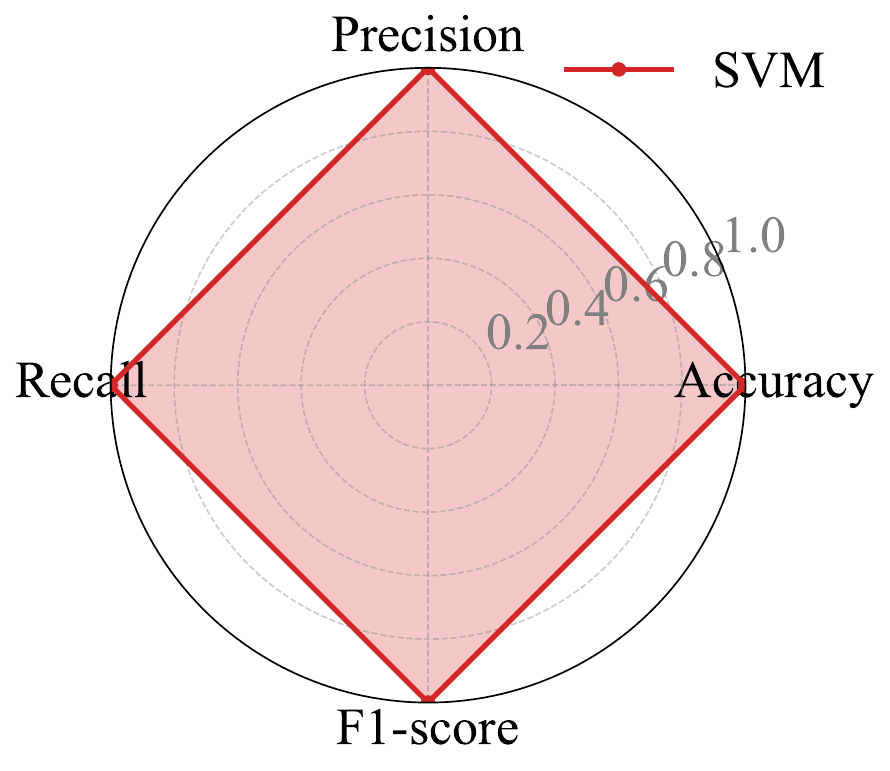}
		\label{fig:subC}}
	\caption{SVM classification performance}
	\label{fig:SVM accuracies}
\end{figure}

We evaluated the classification performance of the SVM on three datasets. As shown in the Figure \ref{fig:SVM accuracies}, the SVM achieved excellent performance in classifying client behaviors.

\begin{table}[htbp]
	\centering
	\normalsize
	\begin{tabular}{l|cc|cc}
		\hline
		& MNIST & FashionMNIST & GTSRB \\
		\hline
		FLTrust   & 26.16\% & 23.62\% & 3.30\% \\
		LFR       & 9.73\%  & 9.99\%  & 97.79\% \\
		Median    & 75.69\% & 63.36\% & 0.42\% \\
		Bulyan    & 9.73\%  & 9.99\%  & 0.42\% \\
		FedGuard  & 96.22\% & 87.87\% & 97.79\% \\
		FedAvg    & 98.56\% & 90.01\% & 99.85\% \\
		\hline
	\end{tabular}
	\caption{Accuracies (\%) of different aggregation rules on IID settings for MNIST, FashionMNIST, and GTSRB.}
	\label{tab:aggregation_results_iid}
\end{table}

\begin{table}[htbp]
	\centering
	\normalsize
	\begin{tabular}{l|cc|cc}
		\hline
		& MNIST & FashionMNIST & GTSRB \\
		\hline
		FLTrust   & 27.42\% & 46.62\% & 0.33\% \\
		LFR       & 9.55\%  & 9.80\%  & 86.66\% \\
		Median    & 64.04\% & 56.07\% & 0.34\% \\
		Bulyan    & 9.55\%  & 9.93\%  & 0.33\% \\
		FedGuard  & 92.93\% & 84.32\% & 87.27\% \\
		FedAvg    & 97.28\% & 85.92\% & 99.18\% \\
		\hline
	\end{tabular}
	\caption{Accuracies (\%) of different aggregation rules on Non-IID settings for MNIST, FashionMNIST, and GTSRB.}
	\label{tab:aggregation_results_non_iid}
\end{table}

\subsection{Experimental Parameter Settings}
\begin{itemize}
	\item \textbf{Dataset settings:} The seed dataset comprises 0.01\% of the entire dataset, and the public dataset is formed by replicating it 100 times. Other robust aggregation strategies that rely on trusted data use a root dataset of the same size as the seed dataset. In the non-IID scenario, the Dirichlet partitioning factor is set to 0.01.
	\item \textbf{Model training settings:} By default, the federated learning setup involves 30 clients and runs for 30 communication rounds. In each round, every client performs 10 local training epochs with a batch size of 32 and a learning rate of 0.01. A total of 100 shadow models are trained for 100 epochs with the same learning rate (0.01). The defense model uses an SVM with an RBF kernel and a penalty coefficient of 1.
	\item \textbf{Attacker settings:} The percentage of Byzantine malicious clients ranges from 10\% to 90\%. When the number of malicious clients is smaller than the number of available attack methods, different attack strategies are randomly assigned. When the number of malicious clients exceeds the number of attack methods, each attack method is assigned to at least one malicious client, and the remaining malicious clients are randomly assigned to available methods. To ensure comparability of poisoning behaviors across different strategies, the same random seed is used in all experiments so that malicious clients perform consistent attacks across methods. For specific attack configurations, the Bit-Flip attack flips the 10th bit of model parameters, while the Lie attack introduces Gaussian noise with an amplification factor of 1. The Random Parameters attack replaces model parameters with uniformly sampled values in the range $[-1, 1]$, and the Krum attack scales local parameters by a factor of 0.5. For data poisoning attacks, 50\% of the local dataset is modified.
\end{itemize}

\begin{figure}[h]
	\centering
	\subfloat[MNIST (IID)]{%
		\includegraphics[width=0.48\linewidth]{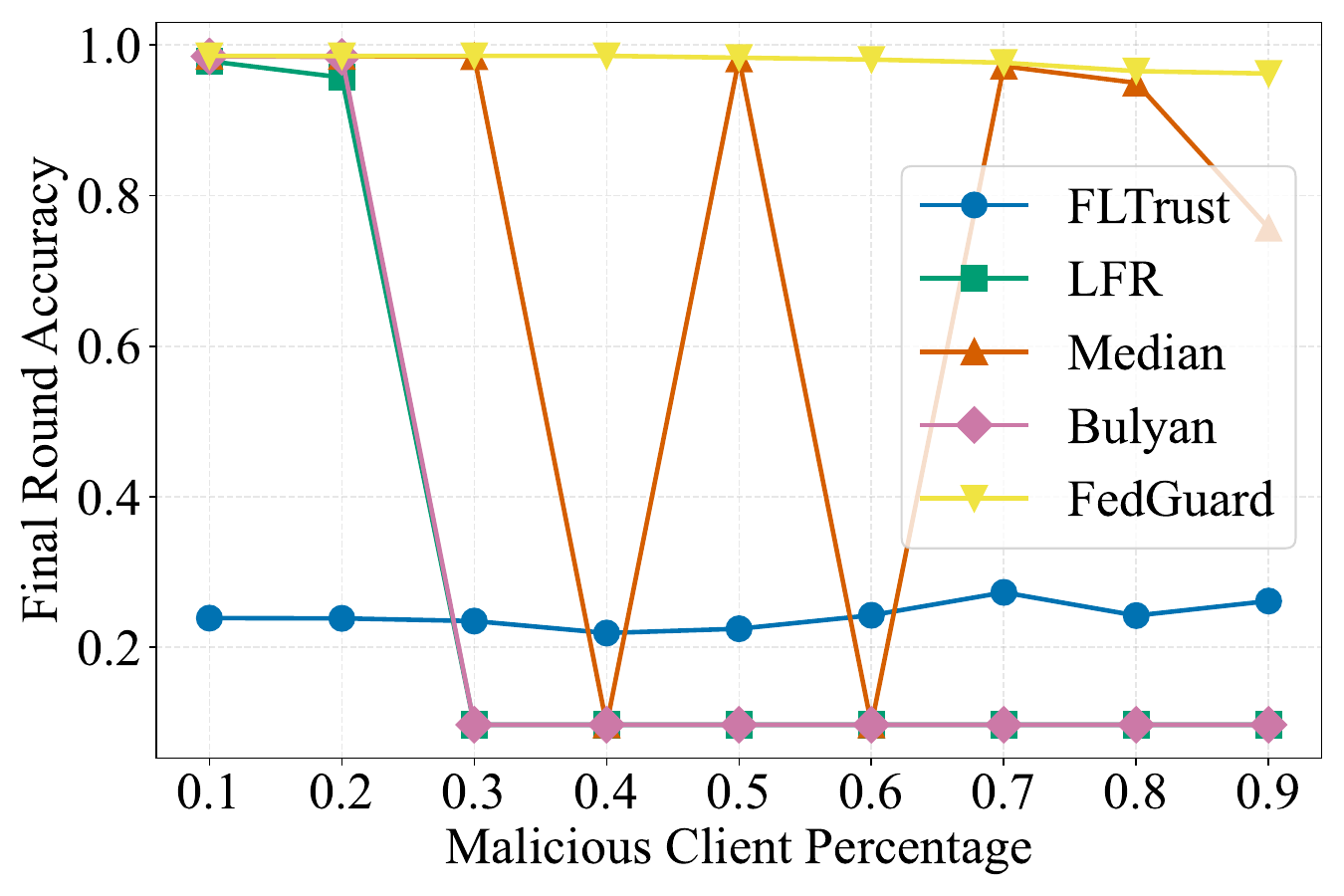}
	}
	\hfill
	\subfloat[FashionMNIST (IID)]{%
		\includegraphics[width=0.48\linewidth]{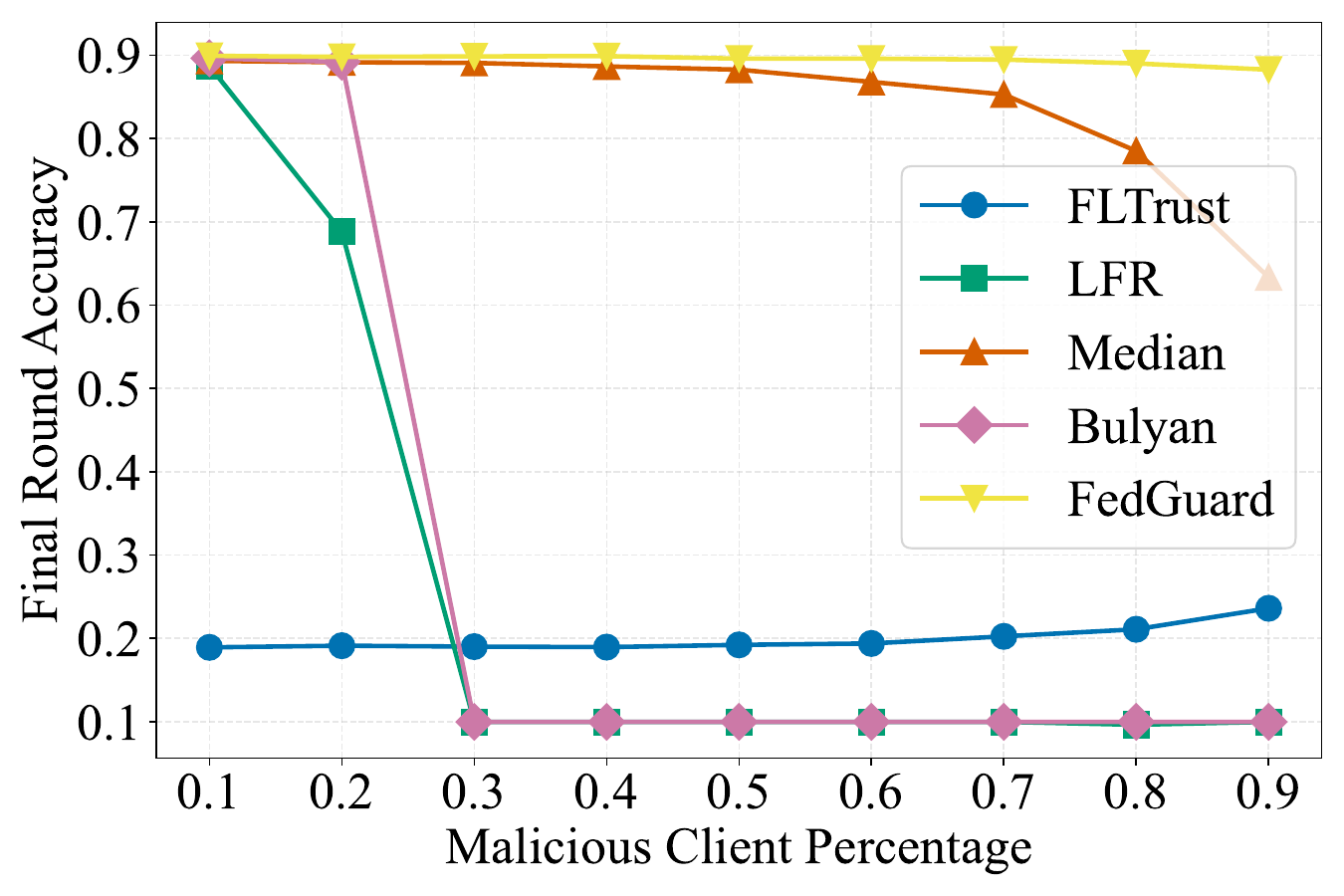}
	}
	\hfill
	\subfloat[GTSRB (IID)]{%
		\includegraphics[width=0.48\linewidth]{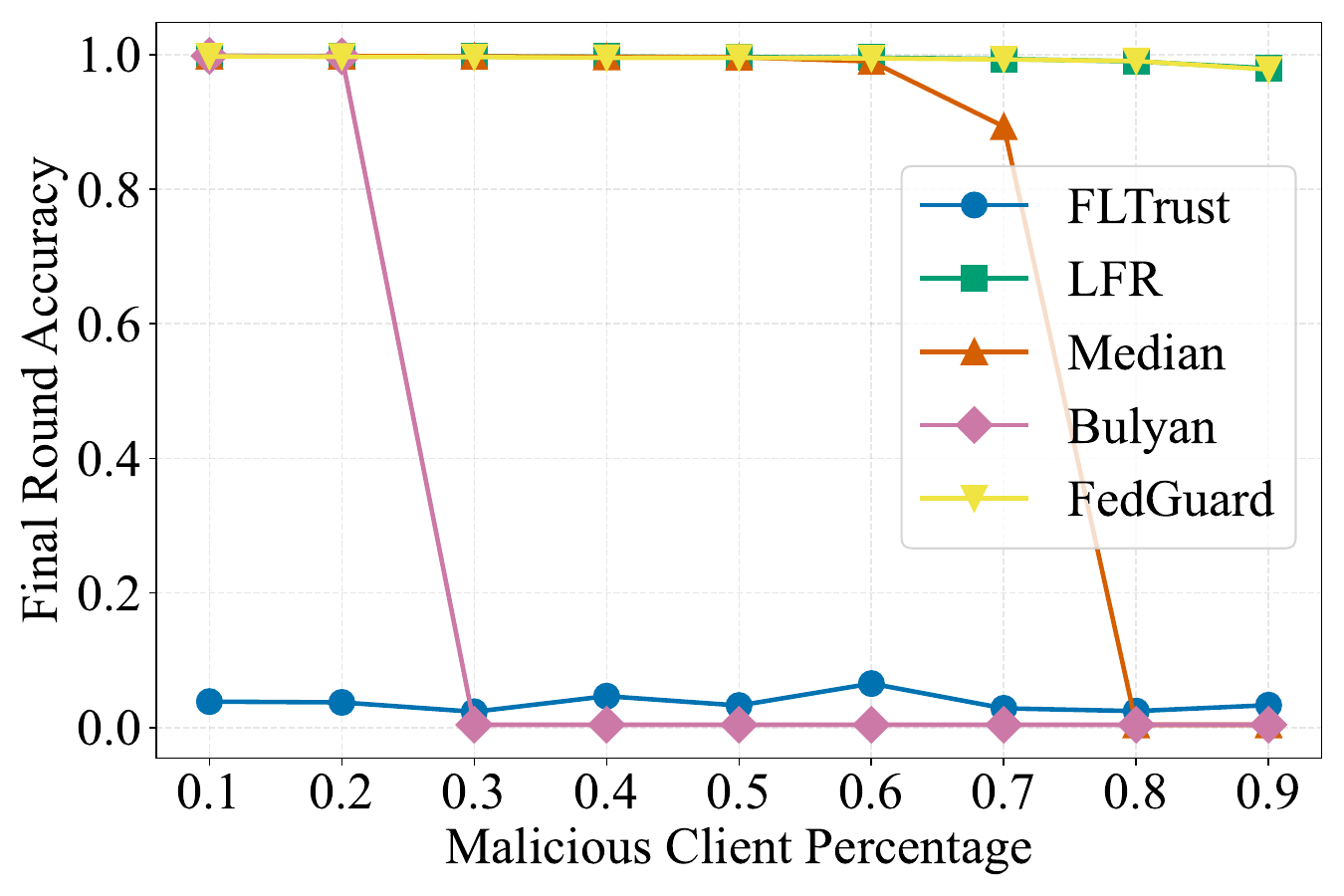}
	}
	\hfill
	\subfloat[MNIST (Non-IID)]{%
		\includegraphics[width=0.48\linewidth]{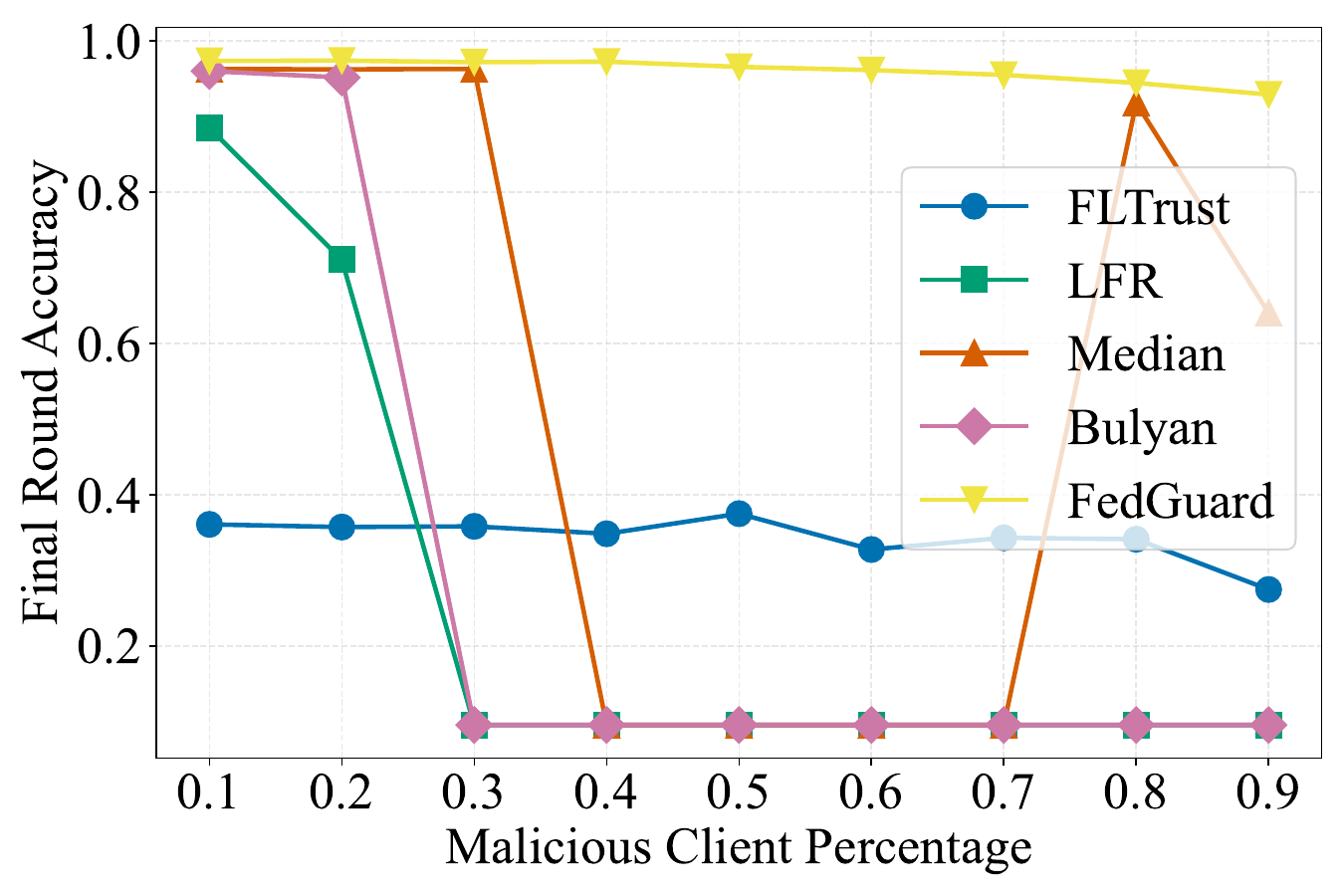}
	}
	\hfill
	\subfloat[FashionMNIST (Non-IID)]{%
		\includegraphics[width=0.48\linewidth]{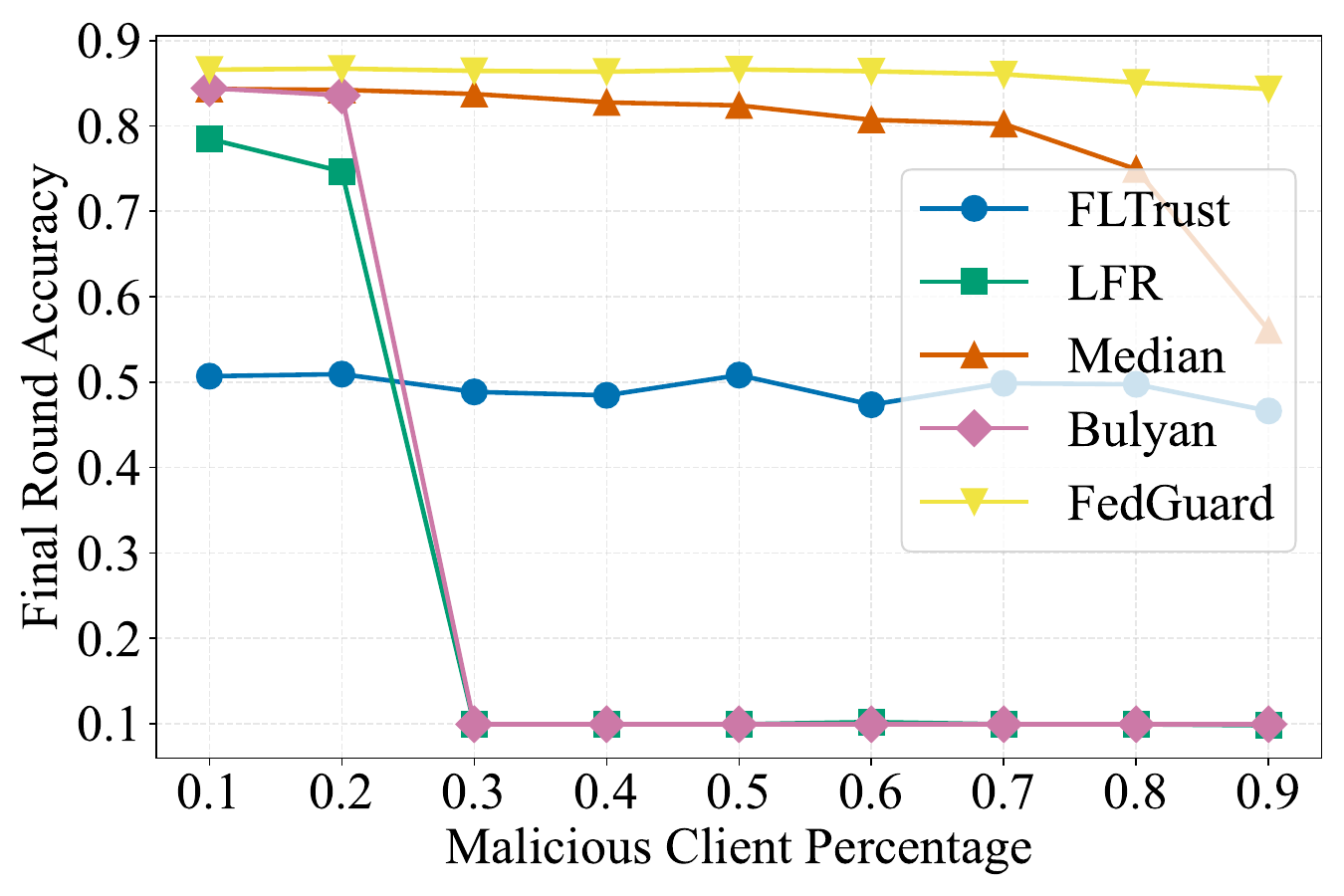}
	}
	\hfill
	\subfloat[GTSRB (Non-IID)]{%
		\includegraphics[width=0.48\linewidth]{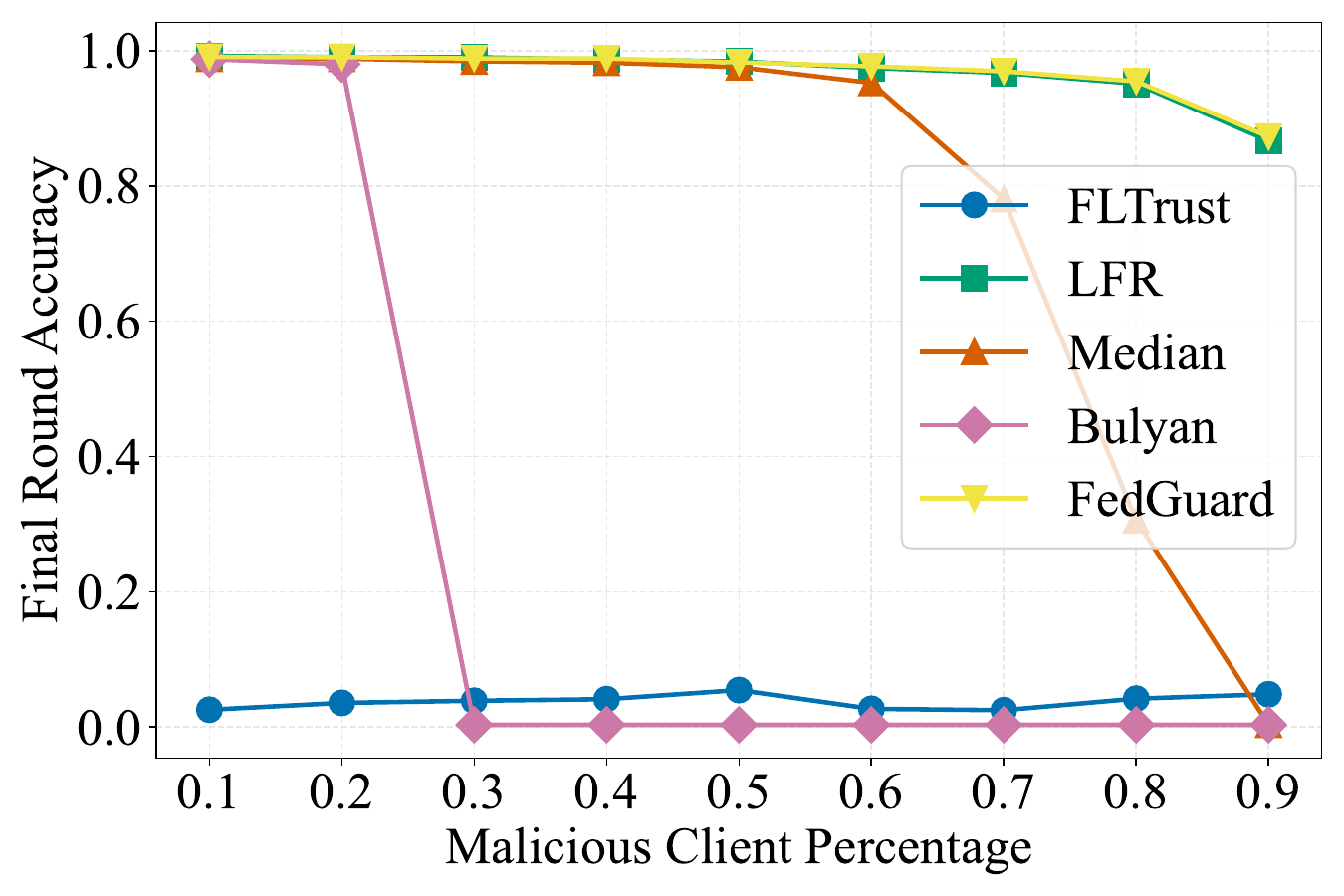}
	}
	\caption{Accuracies versus Byzantine percentage under IID and Non-IID settings.}
	\label{fig:accuracies_vs_percentage}
\end{figure}

\begin{table*}[htbp]
	\centering
	\normalsize
	\begin{tabular}{l|ccccccc|c}
		\hline
		Aggregation Rule & \shortstack{Label-Flipping} & \shortstack{Backdoor \\ Injection} & Sign-Flip & \shortstack{Random \\ Parameters} & Bit-Flip & Lie & Krum & Accuracies \\
		\hline
		FLTrust   & 9 & 0 & 0 & 1 & 0 & 10 & 10 & 57.87\% \\
		LFR       & 0 & 1 & 0 & 0 & 3 & 0 & 1 & 9.53\% \\
		Median    & 10 & 10 & 10 & 10 & 10 & 10 & 10 & 9.53\% \\
		Bulyan    & 0 & 0 & 0 & 0 & 0 & 9 & 1 & 9.53\% \\
		FedGuard  & 0 & 0 & 0 & 0 & 0 & 0 & 0 & 84.82\% \\
		\hline
	\end{tabular}
	\caption{Mistaken selection count and accuracies of different aggregation rules against various attacks.}
	\label{tab:performance_diff_aggs}
\end{table*}

\subsection{Performance Metrics}
\begin{figure}[h]
	\centering
	\subfloat[MNIST (IID)]{%
		\includegraphics[width=0.48\linewidth]{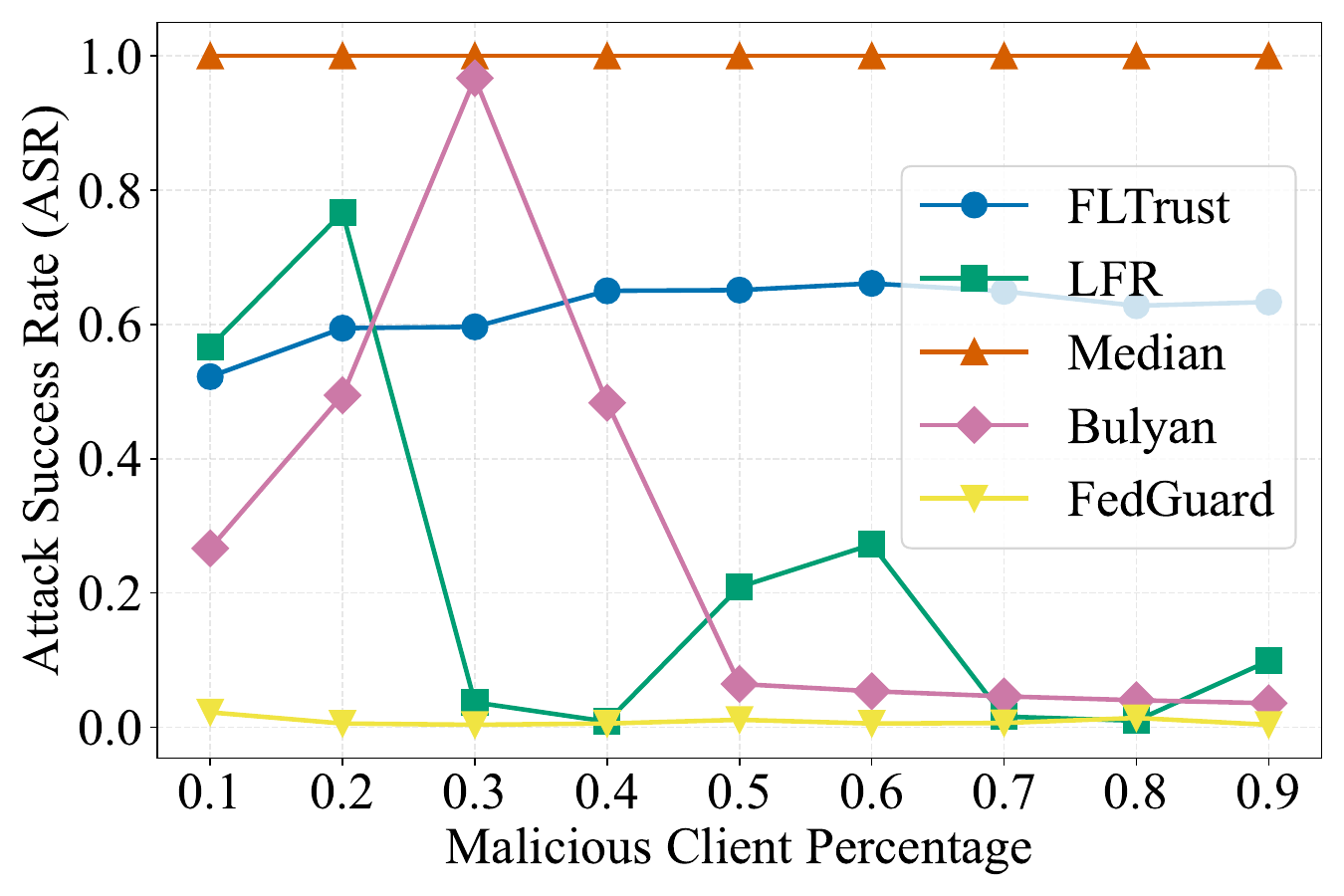}
	}
	\hfill
	\subfloat[FashionMNIST (IID)]{%
		\includegraphics[width=0.48\linewidth]{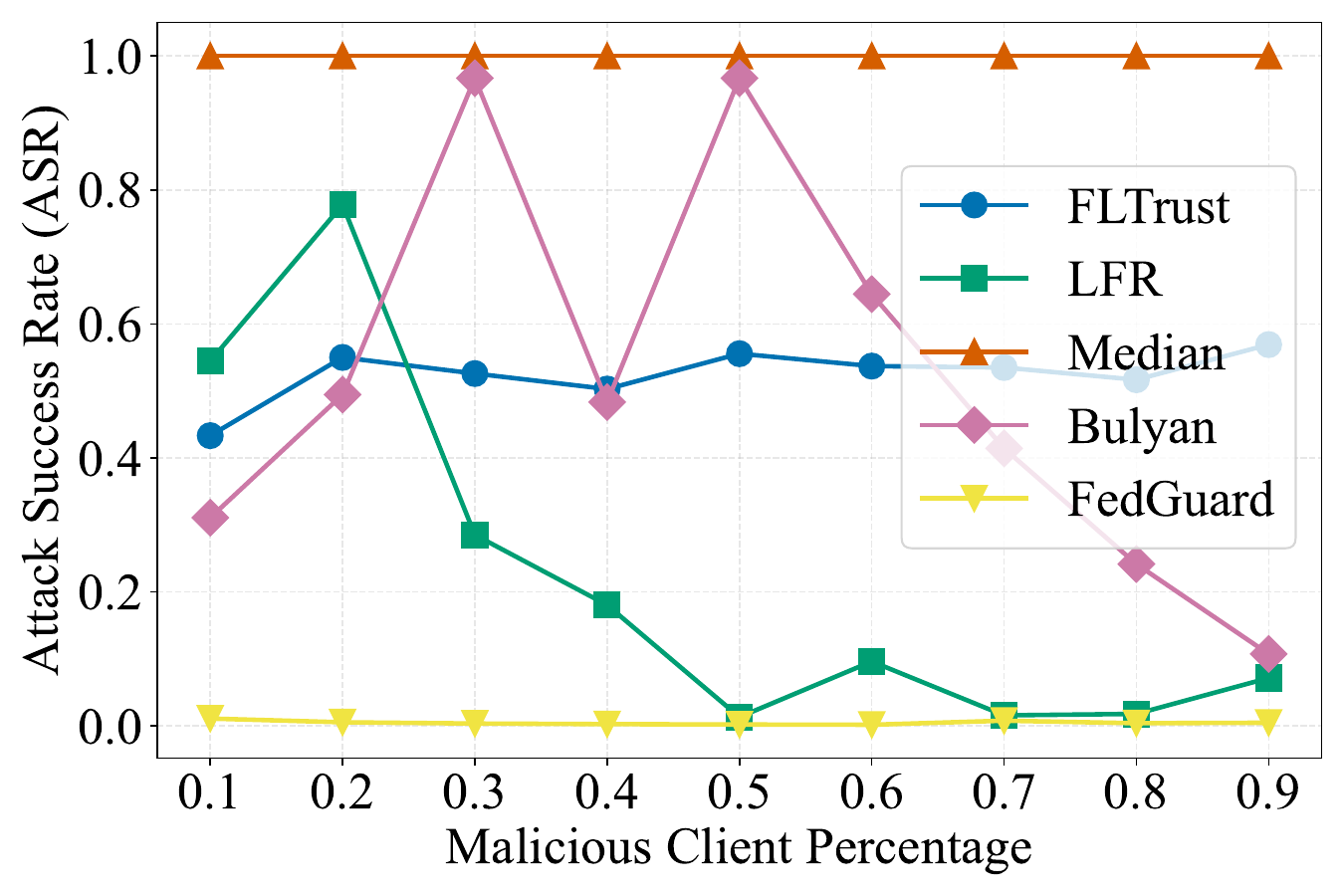}
	}	
	
	\subfloat[GTSRB (IID)]{%
		\includegraphics[width=0.48\linewidth]{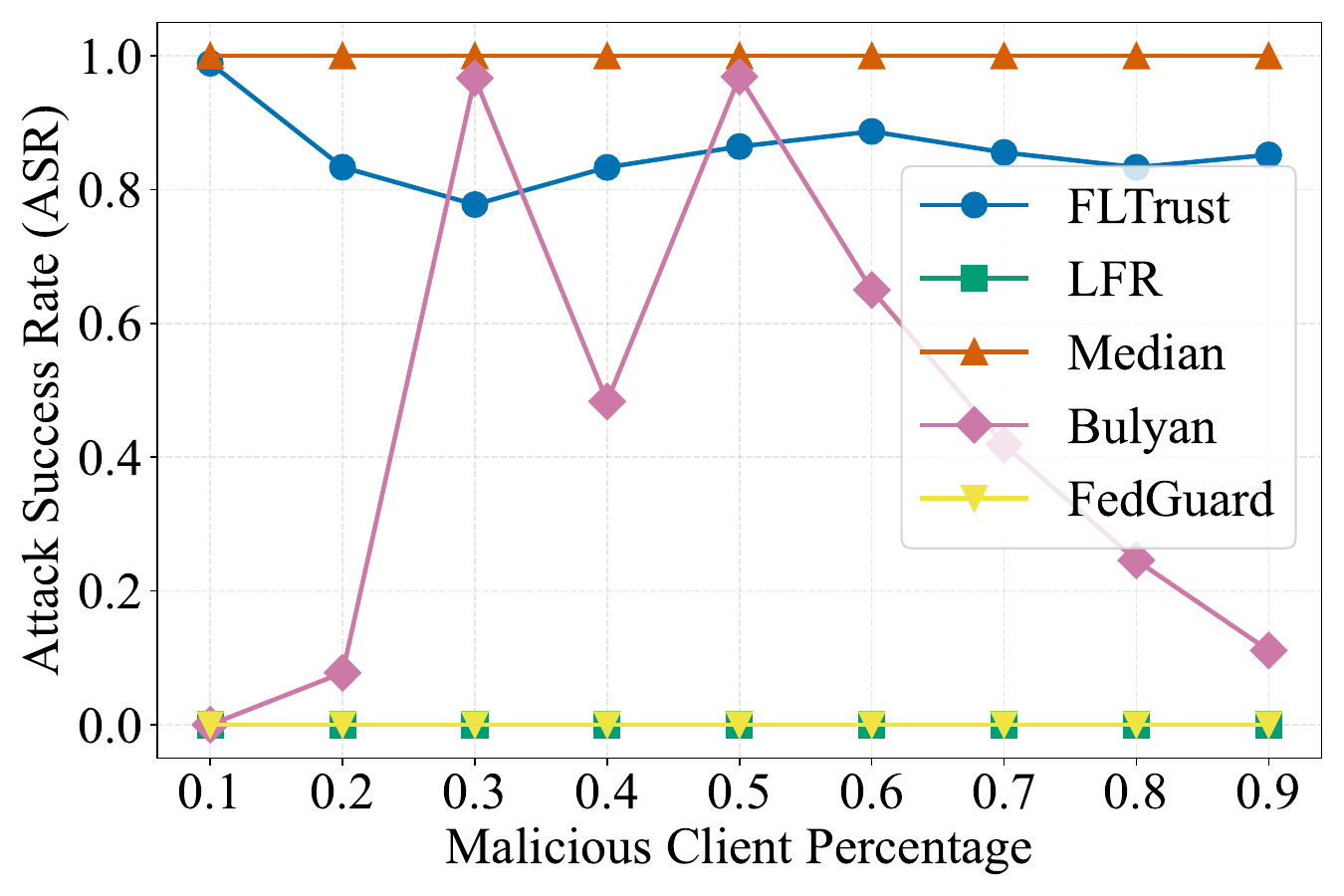}
	}
	\hfill
	\subfloat[MNIST (Non-IID)]{%
		\includegraphics[width=0.48\linewidth]{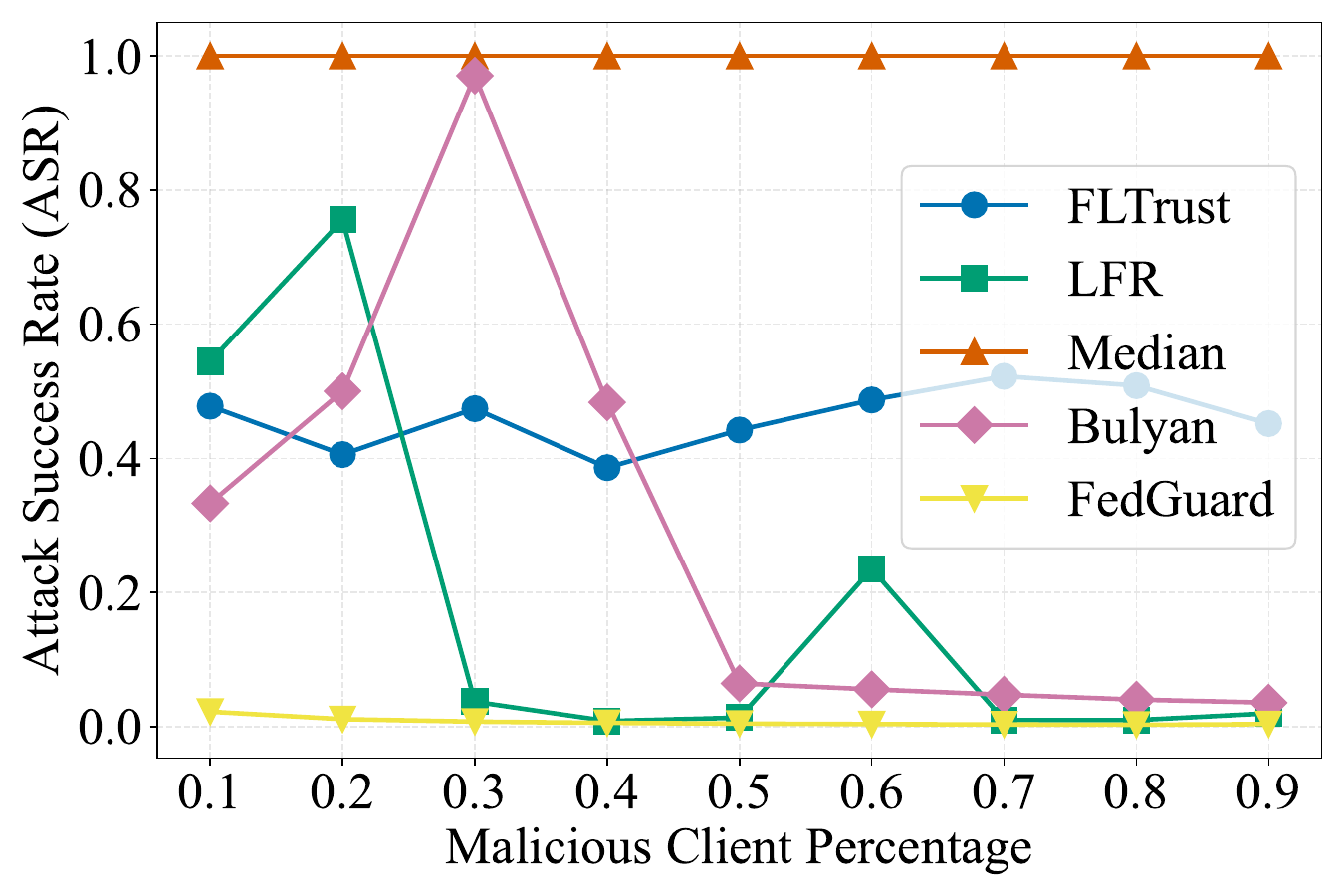}
	}
	
	\subfloat[FashionMNIST (Non-IID)]{%
		\includegraphics[width=0.48\linewidth]{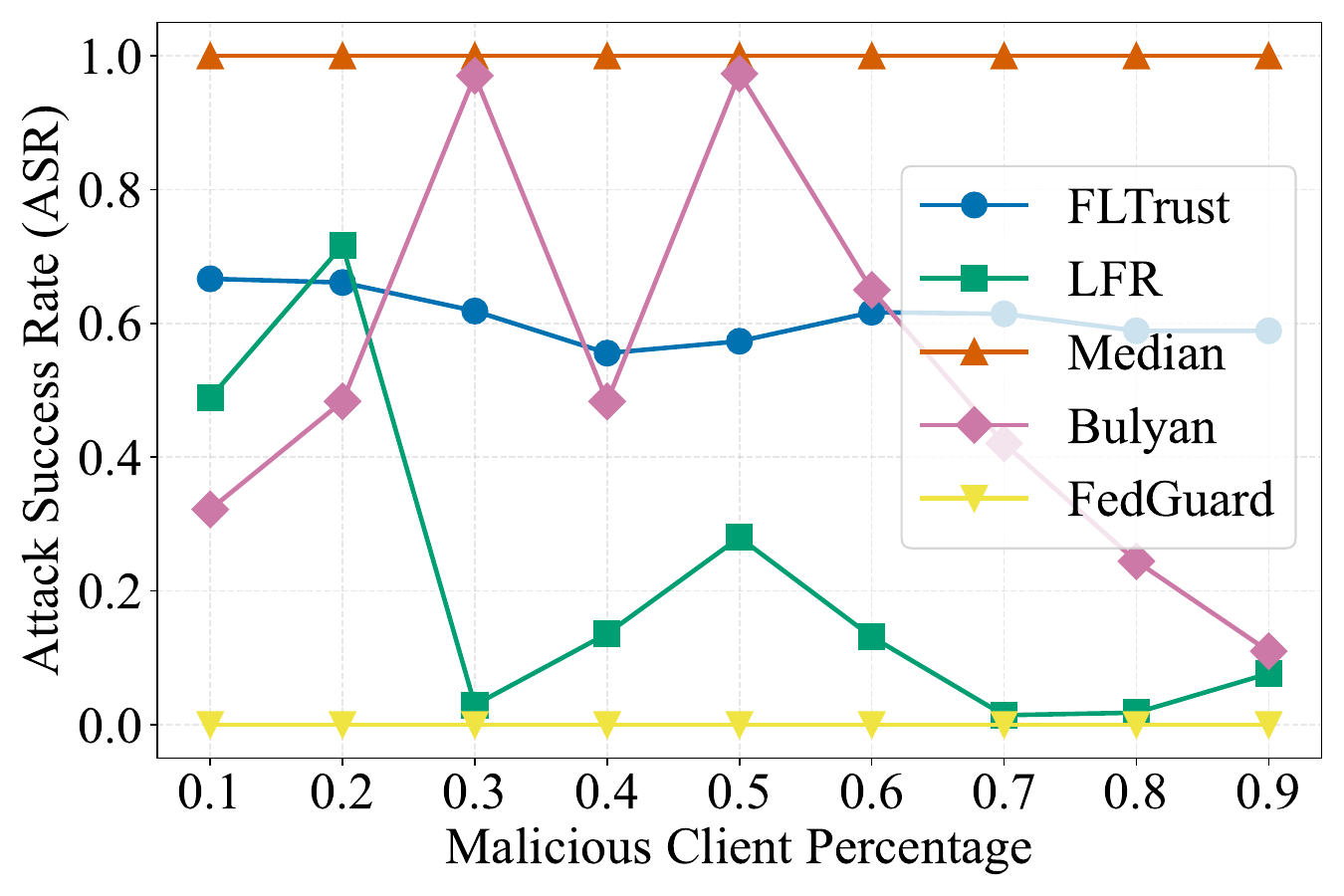}
	}
	\hfill
	\subfloat[GTSRB (Non-IID)]{%
		\includegraphics[width=0.48\linewidth]{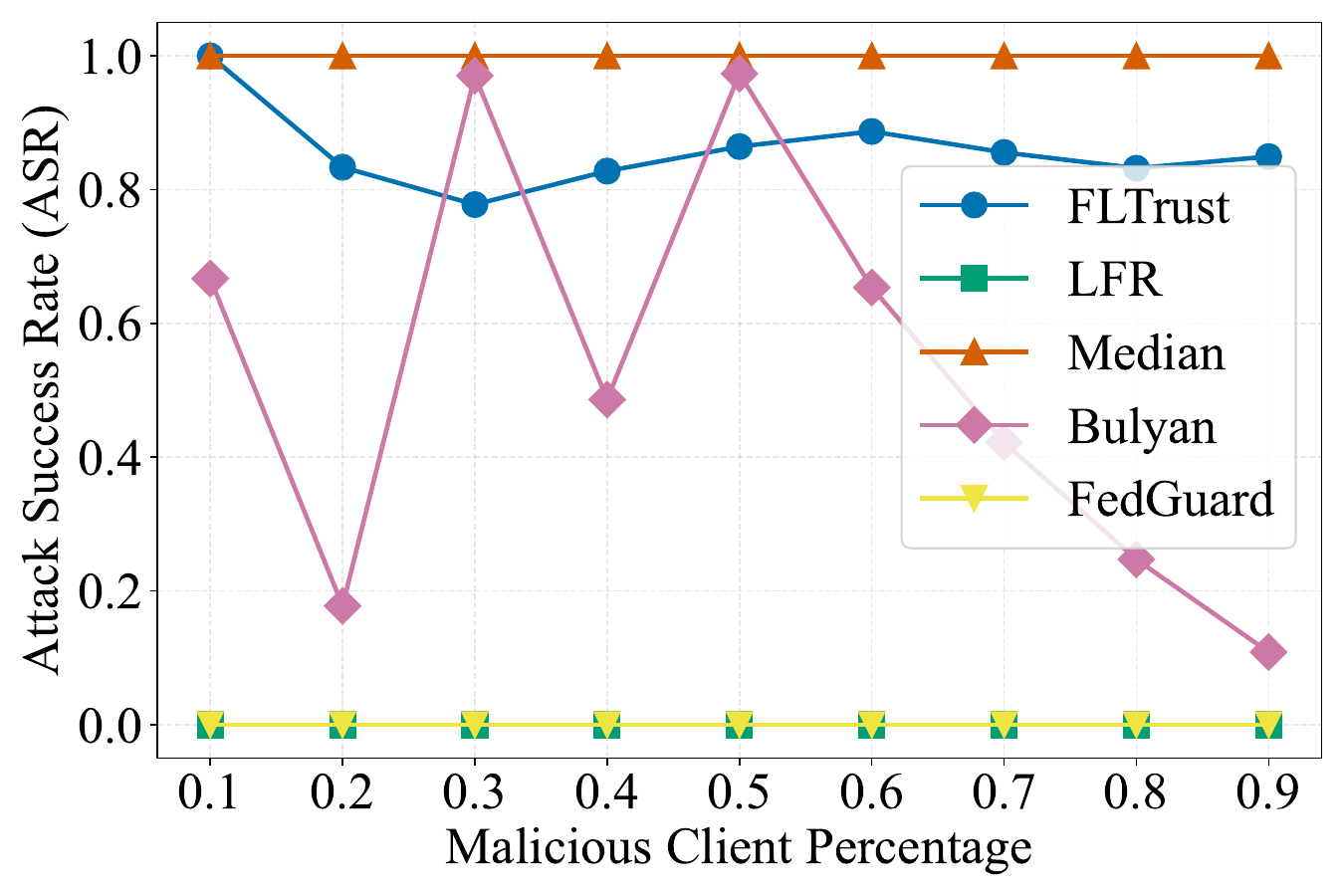}
	}
	\caption{AER versus Byzantine percentage under IID and Non-IID settings.}
	\label{fig:aer_vs_percentage}
\end{figure}

\begin{itemize}
	\item \textbf{Model Accuracy:} We evaluated model performance by comparing accuracy under Byzantine attacks. As shown in Figure \ref{fig:accuracies_vs_percentage}, FedGuard outperforms other algorithms, achieving superior accuracy under various attacks.
	\item \textbf{Average Error Rate (AER):} Unlike traditional error rate calculation methods \cite{ZSW2022}, we adopt a stricter criterion, where each instance of the server mistakenly selecting a Byzantine client is considered a successful attack. Let $N_b$ denote the total number of Byzantine attackers, and $N_m$ the number of Byzantine clients incorrectly identified as benign by the server in each round. The final average Byzantine attack error rate is defined as $\mathrm{AER} = \frac{1}{R} \sum_{r=1}^R \frac{N_m^r}{N_b^r},$ where $r$ represents the current round, and $R$ is the total number of rounds. The results, shown in Figure \ref{fig:aer_vs_percentage}, demonstrate that FedGuard exhibits exceptional Byzantine robustness, significantly outperforming other methods.		
\end{itemize}

\subsection{Convergence Performance of FedGuard}
\begin{figure}[htbp]
	\centering

	\subfloat[MNIST]{%
		\includegraphics[width=0.32\linewidth]{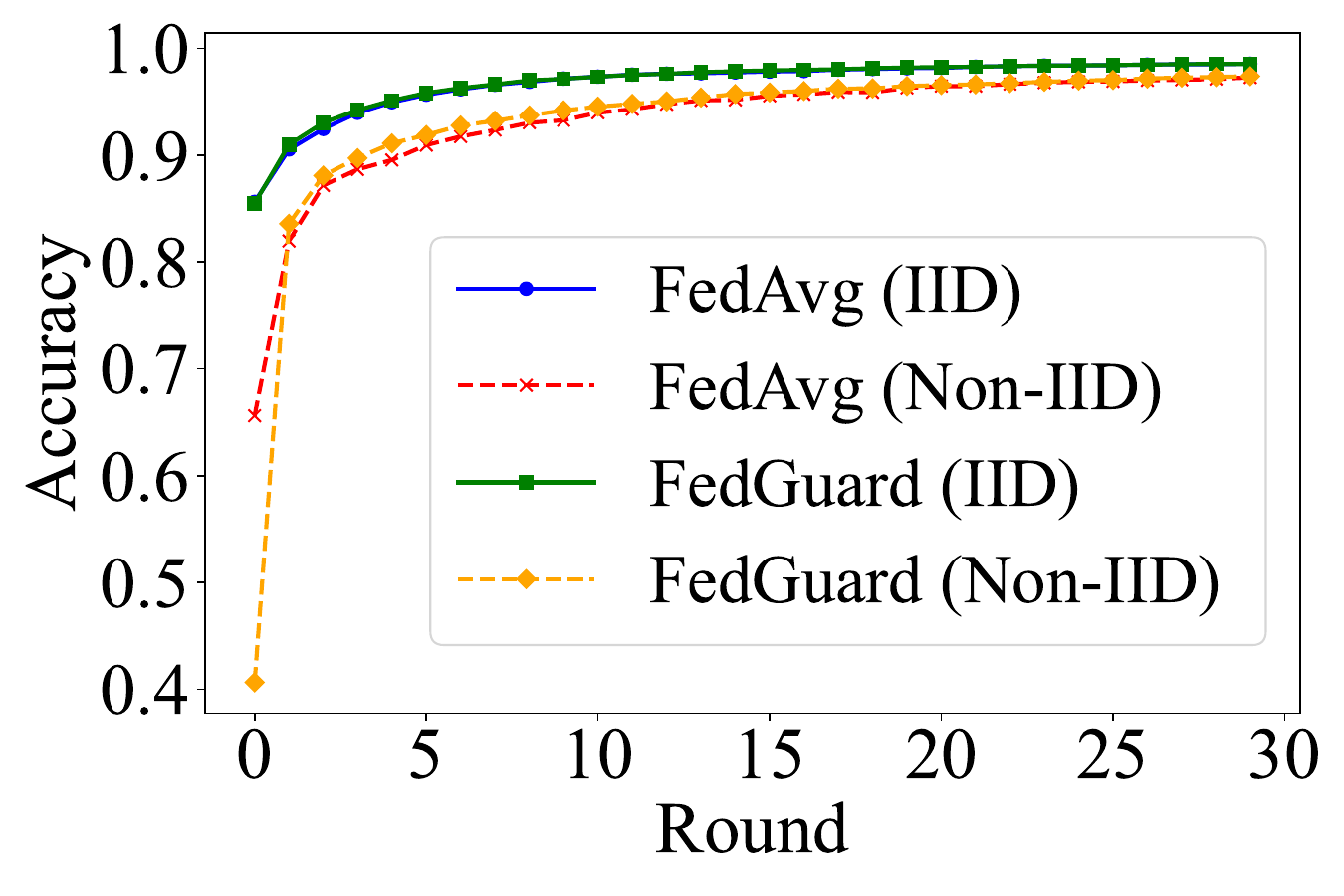}
	}
	\hfill
	\subfloat[FashionMNIST]{%
		\includegraphics[width=0.32\linewidth]{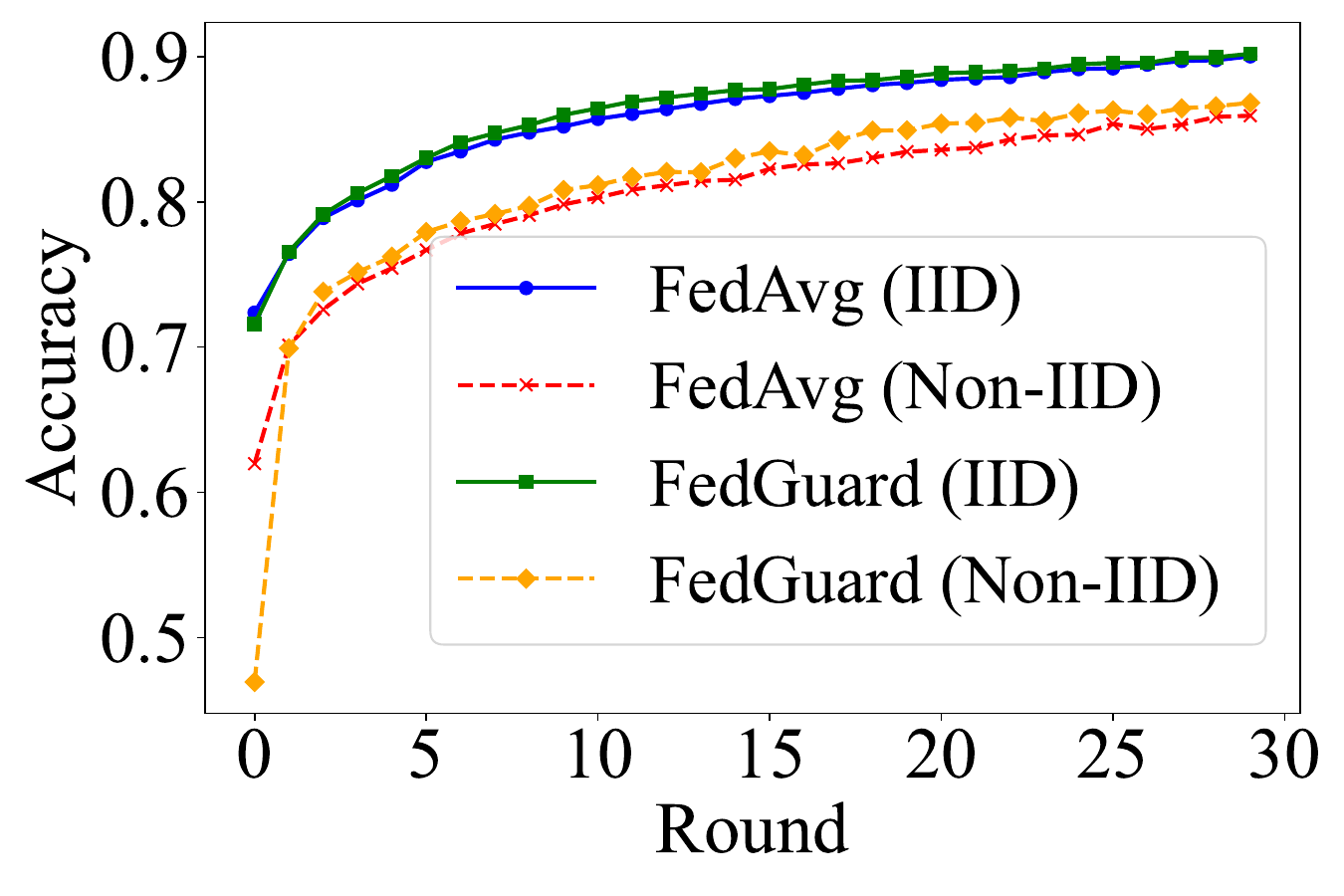}
	}
	\hfill
	\subfloat[GTSRB]{%
		\includegraphics[width=0.32\linewidth]{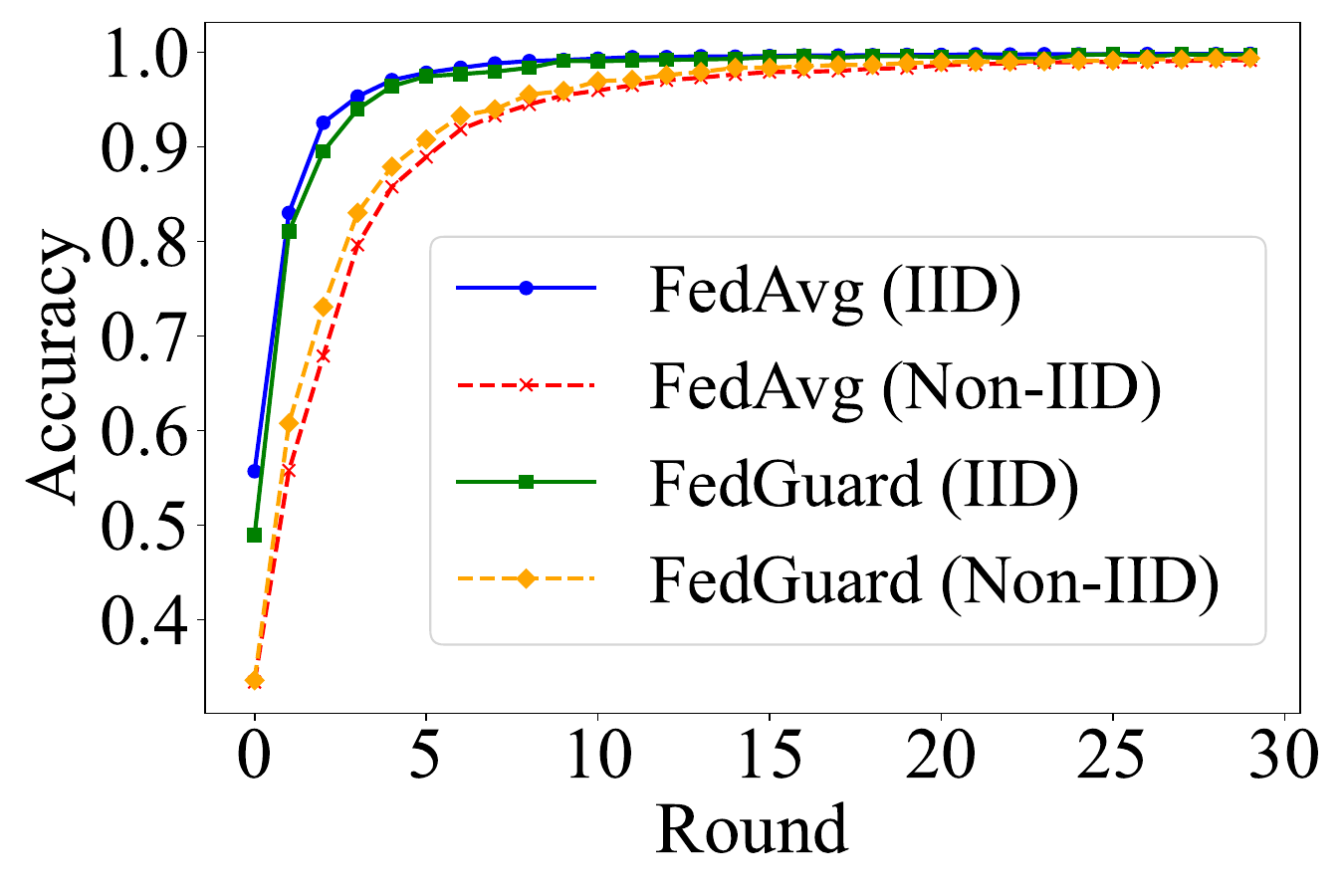}
	}
	
	\caption{Accuracies without Byzantine attacks for MNIST, FashionMNIST, and GTSRB.}
	\label{fig:convergence_comparison}
\end{figure}

We consider an ideal case where all clients are benign. The test accuracy of the global models trained by FedGuard and FedAvg is presented in Figure \ref{fig:convergence_comparison}. The results show that FedGuard achieves the same convergence performance as FedAvg when all clients are benign.

\subsection{Privacy-preserving Membership Inference}
In each round, a privacy-preserving inference attack is conducted on participant models using the public dataset. From the server’s perspective, since each client trains only on the public dataset, this inference attack cannot reveal any additional private information.

\subsection{Model Availability}
We evaluated the availability of the global model under the simultaneous presence of 7 attack types, with 90\% of clients being Byzantine, where FedAvg represents the accuracy without Byzantine attacks. The accuracy comparisons under different aggregation rules are presented in the Tables \ref{tab:aggregation_results_iid} and \ref{tab:aggregation_results_non_iid}. FedGuard consistently performs close to FedAvg, demonstrating its robustness and superiority over other methods.

Specifically, in a more extreme setting on the MNIST dataset, a total of 8 clients were involved, 7 of which were malicious. Each malicious client launched a distinct attack, resulting in 7 different attack types in total. The federated training was conducted for 10 rounds, with a Dirichlet partitioning factor of 0.01, and the seed dataset size was set to 0.01\% of the original dataset. The public dataset is constructed by replicating the seed dataset 200 times. As shown in the Table \ref{tab:performance_diff_aggs}, we report the exact number of times the server mistakenly selected malicious clients under different attack types. The results demonstrate that FedGuard consistently maintains the lowest misclassification rate and the highest accuracy, ensuring model availability even under such an extreme environment.

\section{Conclusion}
This paper proposes FedGuard, a new Byzantine-robust mechanism. It enables distributed collaborative training in highly non-IID environments without compromising model accuracy, even with up to 90\% of clients being malicious and varying types of Byzantine attacks occurring in each training round. FedGuard analyzes client behavior patterns by exploiting membership inference vulnerabilities to identify benign and malicious clients. It integrates this method into the Federated Learning framework using a two-stage approach with offline and online phases. In the offline phase, the server creates shadow models that simulate the behaviors of benign and malicious clients using a small set of clean data. It randomly selects one from the shadow models of simulated benign clients as the reference model. Then, it calculates MSE and TCD according to the outputs of the shadow models and the reference model, and uses these metrics to train a defense model. In the online phase, the server sends the small set of clean data to the online clients, who use this dataset and their private datasets to train local models and upload them. The server uses the small dataset, the reference model, the uploaded local models, and the defense model to identify any malicious clients, excluding them from the aggregation process. Experimental results demonstrate that FedGuard outperforms existing robust FL solutions across various conditions. Future research will investigate advanced neural network architectures for training defense models to achieve fine-grained classification of attack behaviors from malicious clients.

\bibliography{aaai2026}
\end{document}